\documentclass[12pt]{article}

\usepackage{epsfig}

\usepackage{amssymb}

\usepackage{dcolumn}
\usepackage{bm}

\def\a {\epsilon}

\def\O  {{\cal O}}

\def\bW {Br^{W{\to}l\nu}}
\def\os {opposite-sign }

\def\ktf {$k_t$-factorization }
\def\ktfa {$k_t$-factorization approach }
\def\ap#1#2#3   {{\rm Ann. Phys. (NY)}       #1 (#3) #2}
\def\apj#1#2#3  {{\rm Astrophys. J.}         #1 (#3) #2}
\def\apjl#1#2#3 {{\rm Astrophys. J. Lett.}   #1 (#3) #2}
\def\app#1#2#3  {{\rm Acta. Phys. Pol.}      #1 (#3) #2}
\def\chp#1#2#3  {{Chin.\ Phys. }             #1 (#3) #2}
\def\cpc#1#2#3  {{\rm Computer Phys. Comm.}  #1 (#3) #2}
\def\dum#1#2#3  {{~}                         #1 (#3) #2}
\def\epjc#1#2#3 {{\rm Eur. Phys. J. C}       #1 (#3) #2}
\def\err#1#2#3  {{\it Erratum}               #1 (#3) #2}
\def\ib#1#2#3   {{\it ibid.}                 #1 (#3) #2}
\def\jcp#1#2#3  {{\rm J. Comp. Phys.}        #1 (#3) #2}
\def\jmp#1#2#3  {{\rm J. Math. Phys.}        #1 (#3) #2}
\def\jhep#1#2#3 {{\rm JHEP}                  #1 (#3) #2}
\def\ijmp#1#2#3 {{\rm Int. J. Mod. Phys.}    #1 (#3) #2}
\def\jpg#1#2#3  {{\rm J. Phys. G.}           #1 (#3) #2}
\def\mpl#1#2#3  {{\rm Mod. Phys. Lett.}      #1 (#3) #2}
\def\nat#1#2#3  {{\rm Nature (London)}       #1 (#3) #2}
\def\ncim#1#2#3 {{\rm Nuovo Cimento}         #1 (#3) #2}
\def\nca#1#2#3  {{\rm Nuovo Cimento A}       #1 (#3) #2}
\def\ncb#1#2#3  {{\rm Nuovo Cimento B}       #1 (#3) #2}
\def\nim#1#2#3  {{\rm Nucl. Instr. Meth.}    #1 (#3) #2}
\def\njp#1#2#3  {{New J. Phys. }             #1 (#3) #2}
\def\np#1#2#3   {{\rm Nucl. Phys.}           #1 (#3) #2}
\def\npb#1#2#3  {{\rm Nucl. Phys. B}         #1 (#3) #2}
\def\pan#1#2#3  {{\rm Phys. At. Nuclei}      #1 (#3) #2}
\def\pl#1#2#3   {{\rm Phys. Lett.}           #1 (#3) #2}
\def\plb#1#2#3  {{\rm Phys. Lett. B}         #1 (#3) #2}  
\def\prep#1#2#3 {{\rm Phys. Rep.}            #1 (#3) #2}    
\def\prev#1#2#3 {{\rm Phys. Rev.}            #1 (#3) #2}
\def\prc#1#2#3  {{\rm Phys. Rev. C}          #1 (#3) #2}
\def\prd#1#2#3  {{\rm Phys. Rev. D}          #1 (#3) #2}
\def\prev#1#2#3 {{\rm Phys. Rev.}            #1 (#3) #2}
\def\prl#1#2#3  {{\rm Phys. Rev. Lett.}      #1 (#3) #2}
\def\prs#1#2#3  {{\rm Proc. Roy. Soc.}       #1 (#3) #2}
\def\ptp#1#2#3  {{\rm Prog. Theor. Phys.}    #1 (#3) #2}
\def\ps#1#2#3   {{\rm Physica Scripta}       #1 (#3) #2}
\def\rmp#1#2#3  {{\rm Rev. Mod. Phys.}       #1 (#3) #2}
\def\rpp#1#2#3  {{\rm Rep. Prog. Phys.}      #1 (#3) #2}
\def\sjnp#1#2#3 {{\rm Sov. J. Nucl. Phys.}   #1 (#3) #2}
\def\spj#1#2#3  {{\rm Sov. Phys. JETP}       #1 (#3) #2}
\def\spu#1#2#3  {{\rm Sov. Phys.-Usp.}       #1 (#3) #2}
\def\yaf#1#2#3  {{\rm Yad. Fiz.}             #1 (#3) #2}
\def\zp#1#2#3   {{\rm Zeit. Phys.}           #1 (#3) #2}
\def\zpa#1#2#3  {{\rm Zeit. Phys. A}         #1 (#3) #2}
\def\zpc#1#2#3  {{\rm Zeit. Phys. C}         #1 (#3) #2}
\def\etal{{\rm et al.}}

\begin{document}


\titlepage
\title{Associated $W^{\pm}D^{(*)}$ production at the LHC\\
     and prospects to observe double parton interactions}
\maketitle




%
\begin{center}
S.P. Baranov\footnote{baranov@sci.lebedev.ru}\\

{\it P.N. Lebedev Institute of Physics, \\
              Lenin Avenue 53, 119991 Moscow, Russia}\\
\vspace{7mm}
A.\ V.\ Lipatov\footnote{lipatov@theory.sinp.msu.ru}, 
%
M.\ A.\ Malyshev\footnote{malyshev@theory.sinp.msu.ru}, 
%
A.\ M.\ Snigirev\footnote{snigirev@lav01.sinp.msu.ru}, 
%
N.\ P.\ Zotov\footnote{zotov@theory.sinp.msu.ru}\\
{\it Skobeltsyn Institute of Nuclear Physics,\\
      Lomonosov Moscow State University, 119991 Moscow, Russia}
\end{center}

%
\begin{abstract}
Associated production of charged gauge bosons and charged charm\-ed mesons
at the LHC is considered in the framework of \ktfa. Theoretical predictions
are compared with ATLAS data, and reasonably good agreement is found.
Predictions on the same-sign $W^{\pm}D^{\pm}$ configurations are presented 
including single parton scattering and double parton scattering contributions. 
The latter are shown to dominate over the former, thus giving evidence that 
the proposed process can serve as another indicator of double parton 
interactions.
\end{abstract}

\begin{flushleft}
\end{flushleft}


\section{Introduction}
Having the LHC put into operation, one got access to a number of 
`rare' processes which would have never been systematically studied 
at the accelerators of previous generations. In this article we draw 
attention to the associated production of weak gauge bosons and 
charmed mesons. This process is interesting on its own as providing 
a complex test of perturbative QCD and our knowledge of parton 
distributions. Moreover, we argue that it can serve as an indicator 
of double parton interactions, nowadays widely discussed in the 
literature \cite{Bartalini,Abramowicz,Bansal}.

This article was greatly stimulated by the recent measurement of the $WD$ 
production cross sections reported by ATLAS Collaboration \cite{ATLAS}. 
In that study, the interest was mainly focused on the properties of 
strange sea (see discussion below) and, therefore, in order to suppress 
other possible contributions (considered in this context as background), 
the authors have only presented the difference between the opposite-sign 
(OS) and same-sign (SS) $WD$ production cross sections, 
$\sigma^{OS-SS}(WD)$. In particular, this excludes the Double Parton 
Scattering (DPS) processes which yield same-sign $W^{\pm}D^{(*)\pm}$ 
and opposite-sign $W^{\pm}D^{(*)\mp}$ combinations with equal 
probability.

On the contrary, we are more interested in just detecting the DPS 
events, and so, will lay emphasis on the SS states. Our article is 
organized as follows. First, we describe our theoretical approach and 
check its validity by comparing with ATLAS data on $\sigma^{OS-SS}(WD)$.
Then we extend our consideration to the SS states and make predictions 
for the production cross sections and  
kinematic observables which could be useful in discriminating the SPS and 
DPS contributions.

\section{Theoretical framework.\\ Comparison with ATLAS data.}

At the parton level, the production of \os $W^{\pm}D^{(*)\mp}$ states 
is dominated by the quark-gluon scattering
\begin{equation}\label{gq2wc}
g+q\to W^-+c \mbox{~~~~or~~~~} g+\bar{q}\to W^++\bar{c}
\end{equation}
followed by nonperturbative fragmentation of $c$-quarks into charmed 
mesons. Here the main contribution comes from strange quarks, while the 
contribution from $d$-quarks is suppressed by Cabibbo angle.

To calculate the physical cross sections, we employ the \ktf approach
\cite{GLR,smallX}.
Here we see certain advantages in the fact that, even with the leading-order 
(LO) matrix elements for hard subprocess, we can include a large piece of 
next-to-leading order (NLO) corrections taking them into account in the form 
of $k_t$-dependent parton densities. In this way we automatically incorporate 
the initial state radiation effects, which play important role in the event 
kinematics. 
Further on, the formally NLO subprocess
\begin{equation}\label{gg2wcs}
g+g\to W^-{+}c{+}\bar{s} \mbox{~~or~~} g+g\to W^+{+}\bar{c}{+}s
\end{equation}
no longer needs to be added because it is aleady contained in (1). Indeed, 
the quark-gluon coupling in subprocess (2) can also be regarded as part of 
the evolution of sea quark densities $q(x)$ and $\bar{q}(x)$ in (1).

On the technical side, our calculations follow standard QCD and electroweak 
theory Feynman rules, but the initial gluon spin density matrix is taken in 
the form \cite{GLR,smallX}
$\overline{\a_g^{\mu}\a_g^{*\nu}} = k_{T}^\mu k_{T}^\nu/|k_{T}|^2,$
where $k_{T}$ is the component of the gluon momentum perpendicular  
to the beam axis. In the collinear limit, when $k_{T}\to 0$, this
expression converges to the ordinary
$\overline{\a_g^{\mu}\a_g^{*\nu}}=-\frac{1}{2}\;g^{\mu\nu}$, while in  
the case of off-shell gluons it contains an admixture of longitudinal
polarization.

In our numerical analysis we used KMR \cite{KMR} parametrization for 
unintegrated quark and gluon distributions with MSTW \cite{MSTW} collinear
densities taken as input;
we used running strong and electroweak coupling constants normalized to
$\alpha_s(m^2_Z){=}0.118$; $\alpha(m^2_Z){=}1/128$; $\sin^2\Theta_W=0.2312$;
the factorization and renormalization scales were chosen as
$\mu^2_R$=$\mu^2_F$=$m^2_T(W)$ $\equiv$ $m^2_W{+}p^2_T(W)$; 
the $c$-quark mass was set to $m_c{=}1.5$~GeV;
$c$-quarks were converted into $D^{(*)}$ mesons using Peterson fragmentation 
function \cite{Peterson} with $\epsilon=0.06$ and normalized to 
$f(c\to D)=0.268$ and $f(c\to D^*) = 0.229$ \cite{JKLZ}.

Our results obtained with this parameter setting are summarized in Table 1,
where we present the $W^{\pm}D^{(*)\mp}$ production cross sections integrated 
over the fiducial phase space region described in Ref. \cite{ATLAS}. 
We observe reasonable agreement with ATLAS data.

\begin{table}
\caption{Measured and predicted cross sections times the $W{\to}l\nu$ 
branching ratio (in pb) integrated over the fiducial region $p_T(l)>20$ GeV, 
$|\eta(l)|<2.5$, $p_T(\nu)>25$ GeV, $p_T(D^{(*)})>8$ GeV, $|\eta(D^{(*)})|<2.2$.}

\begin{center}
\begin{tabular}{|l|c|c|}
\hline
& &\\
~  & Data & Theory  \\ \hline
& &\\
$Br^{W{\to}l\nu}\sigma^{OS-SS}(W^+D^-)$     & 17.8 &  17.7  \\
& &\\
$Br^{W{\to}l\nu}\sigma^{OS-SS}(W^-D^+)$     & 22.4 &  19.5  \\
& &\\
$Br^{W{\to}l\nu}\sigma^{OS-SS}(W^+D^{*-})$  & 21.2 &  15.1  \\
& &\\
$Br^{W{\to}l\nu}\sigma^{OS-SS}(W^-D^{*+})$  & 22.1 &  16.8  \\
& &\\
\hline
\end{tabular} 
\end{center}
\end{table}

\section{Same-sign $W^{\pm}D^{\pm}$ states\\ and double parton interactions}

Now, having our approach validated, we turn to double parton scattering. 
Detecting same-sign $W^{\pm}D^{(*)\pm}$ configurations is certainly preferable 
here, because we are then free from SPS background due to subprocesses (1) or 
(2). There are, however, still many other background sources, both direct and 
indirect. Among the direct ones, we consider the quark-antiquark annihilation 
at ${\O}(\alpha_s^2\alpha)$

\begin{equation}\label{qq2wcc}
u+\bar{d}\to W^+{+}c{+}\bar{c} \mbox{~~or~~} d+\bar{u}\to W^-{+}c{+}\bar{c}
\end{equation}
and quark-gluon scatering at ${\O}(\alpha_s^3\alpha)$
\begin{equation}\label{gq2qwcc}
g{+}u\to W^+{+}d{+}c{+}\bar{c} \mbox{~~or~~} g{+}d\to W^-{+}u{+}c{+}\bar{c}.
\end{equation}
Subprocess (\ref{gq2qwcc}) has one extra $\alpha_s$ in comparison with
(\ref{qq2wcc}), but it employs gluons which are more abundant than
antiquarks in the proton, and that is why may take over.
%
Among the indirect sources we have gluon-gluon fusion
\begin{equation}\label{gg2wcb}
g+g\to W^-{+}c{+}\bar{b} \mbox{~~or~~} g+g\to W^+{+}b{+}\bar{c}
\end{equation}
followed by the decays $b\to c{+}X$ or $\bar{b}\to\bar{c}{+}X$, and the 
production of top quark pairs
\begin{equation}\label{gg2tt}
g+g\to t+\bar{t} \mbox{~~and~~} q+\bar{q}\to t+\bar{t}
\end{equation}
followed by a long chain of decays, such as $t\to W^+{+}b$, 
$W^+\to c{+}\bar{s}$, $b\to c{+}X$ or $b\to c{+}\bar{c}{+}s$ (and the charge
conjugated modes). Here same-sign $W^{+}D^{(*)+}$ configurations may be formed
by a $W^+$ boson coming from $t$ and a $c$-quark coming from $b$ coming from 
$t$, or a $c$-quark coming from $\bar{b}$ coming from $\bar{t}$.
Similarly, in the case of single top production
\begin{equation}\label{qq2tb}
u+\bar{d}\to t+\bar{b} \mbox{~~or~~} d+\bar{u}\to \bar{t} + b
\end{equation}
the $W^{+}D^{(*)+}$ configuration may be formed by a $W^+$ boson coming from 
$t$ and a $c$-quark coming from $b$ coming from the same $t$, or a $c$-quark 
coming from $\bar{b}$.
Note that the subprocesses (\ref{gg2tt}) are purely strong, 
and so, may have large cross sections in spite of large $t$-quark mass. 
All other possible processes beyond (\ref{qq2wcc})-(\ref{qq2tb}) are expected 
to be suppressed 
by extra powers of coupling constants (already the case of (\ref{qq2tb})) 
or by Kobayashi-Maskawa mixing matrix (already the case of (\ref{gg2wcb})).

A comment is needed on the choice of renormalization scale in (\ref{qq2wcc}).
This process factorises into the production of $W{+}g^*$ 
at $\mu_R^2{=}m^2_T(W)$ and the subsequent gluon splitting $g^{*}{\to}c\bar{c}$,
for which the $c\bar{c}$ invariant mass seems to be a more suitable measure.
Note that using different $\alpha_s$ values for these two different steps does 
not violate the overall gauge invariance. So, we calculate the resulting cross 
section with
$\alpha_s(m^2_T(W))\alpha_s(m^2_{c\bar{c}})$, regarding it as the pessimistic
(the upper) limit for the background.
By the same reasoning, we adopt 
$\alpha^2_s(m^2_T(W))\alpha_s(m^2_{c\bar{c}})$ for subprocess (\ref{gq2qwcc}).

For the fiducial phase space of Ref. \cite{ATLAS}, we estimate the above
contributions to $W^+D^+$ and $W^-D^-$ states as
\begin{eqnarray}
&&\bW\sigma^{W^+D^+}(u\bar{d}{\to}Wc\bar{c})= 0.41 \mbox{~pb}; \label{s1} \\
&&\bW\sigma^{W^-D^-}(d\bar{u}{\to}Wc\bar{c})= 0.29 \mbox{~pb}; \label{s2} \\
&&\bW\sigma^{W^+D^+}(gu{\to}Wdc\bar{c})= 1.0 \mbox{~pb};      \label{s1q}\\
&&\bW\sigma^{W^-D^-}(gd{\to}Wuc\bar{c})= 0.7 \mbox{~pb};      \label{s2q}\\
&&\bW\sigma^{W^+D^+}(gg{\to}Wb\bar{c})= 0.002 \mbox{~pb};     \label{s3p}\\
&&\bW\sigma^{W^-D^-}(gg{\to}Wb\bar{c})= 0.002 \mbox{~pb};     \label{s3n}\\
&&\bW\sigma^{W^+D^+}(gg{\to}t\bar{t})= 1.1 \mbox{~pb};        \label{s4p}\\
&&\bW\sigma^{W^-D^-}(gg{\to}t\bar{t})= 1.1 \mbox{~pb};        \label{s4n}\\
&&\bW\sigma^{W^+D^+}(q\bar{q}{\to}t\bar{t})= 0.6 \mbox{~pb};  \label{s5p}\\
&&\bW\sigma^{W^-D^-}(q\bar{q}{\to}t\bar{t})= 0.6 \mbox{~pb};  \label{s5n}\\
&&\bW\sigma^{W^+D^+}(u\bar{d}{\to}t\bar{b})= 0.06 \mbox{~pb}; \label{s6} \\
&&\bW\sigma^{W^-D^-}(d\bar{u}{\to}b\bar{t})= 0.04 \mbox{~pb}. \label{s7}
\end{eqnarray}

The results (\ref{s1})-(\ref{s2q}) were obtained assuming the already 
mentioned fragmentation probability $f(c\to D)=0.268$. 
The results (\ref{s3p})-(\ref{s7}) were obtained under
the assumption of $100\%$ branching fraction for $t\to bW$, 
of equal fragmentation probabilities for $b\to\bar{B}^0$ and $b\to B^-$, 
and using the inclusive branching fractions 
$Br(\bar{B}^0\to D^+X)=37\%$, $Br(B^0\to D^+X)=3\%$,
$Br(B^-\to D^+X)=10\%$, $Br(B^+\to D^+X)=2.5\%$
listed in the Particle Data Book \cite{PDG}.
The quark masses were set to  $m_t{=}175$~GeV and $m_b{=}4.8$~GeV.
We make no predictions for $W^{\pm}D^{*\pm}$ states for the reason of not 
knowing the relevant $B\to D^*X$ decay branchings.
Variations in $\mu_R^2$ and $\mu_F^2$ within a factor of 2 around the
default value make a factor of 1.6 increasing or decreasing effect on
the estimated production rate.
However, these effects mostly cancel out in the signal to background ratio.

Now we proceed to discussing the expected signal from double parton 
interactions.
Under the hypothesis of having two independent hard partonic subprocesses
$A$ and $B$ in a single $pp$ collision, and under further assumption that
the longitudinal and transverse components of generalized parton
distributions factorize from each other, the inclusive DPS cross section
reads
(for details see, e.g., the recent review \cite{Bartalini} and references 
therein)
\begin{equation} \label{doubleAB}
\sigma^{\rm AB}_{\rm DPS} = \frac{\displaystyle{m}}{\displaystyle{2}} 
\frac{\displaystyle{\sigma^{ A}_{\rm SPS}\sigma^{ B}_{\rm SPS}}}%
{\displaystyle{\sigma_{\rm eff}}},
\end{equation}
where $\sigma_{\rm eff}$ is a normalization cross section that encodes all
``DPS unknowns'' into a single parameter which can be experimentally mesured.
One can identify $\sigma_{\rm eff}$  with the inverse of the proton overlap 
functions squared:
\begin{equation}
\sigma_{\rm eff}=\Bigl[ \int d^2b\,\bigl( T({\bf b}) \bigr)^2 \Bigr]^{-1},
\end{equation}
where $T({\bf b}) = \int f({\bf b_1}) f({\bf b_1{-}b})\,d^2b_1 $ is the
overlap function that characterizes the transverse area occupied by the 
interacting partons, and $f({\bf b})$ is supposed to be a universal
function of the impact parameter ${\bf b}$ for all kinds of partons with 
its normalization fixed as
\begin{eqnarray}
\label{f}
\int f({\bf b_1}) f({\bf b_1{-}b})\,d^2b_1\,d^2b=\int T({\bf b})\,d^2b=1.
\end{eqnarray}
A numerical value of $\sigma_{\rm eff}\simeq$ 15 mb has been obtained 
empirically from fits to $p\bar{p}$ and $pp$ data \cite{cdf4j,cdf,d0,atlas,cms}.
It will be used in our further analysis, although an estimate as low as 
$\sigma_{\rm eff}\simeq$ 5 mb is also present in \cite{D0}.

The inclusive SPS cross sections $\sigma^{A}_{\rm SPS}$ and
$\sigma^{B}_{\rm SPS}$ for the individual partonic subrocesses $A$ and  
$B$ can be calculated in a usual way using the ordinary parton distribution
functions. The symmetry factor $m$ equals to 1 for identical subprocesses
and 2 for the differing ones. 

In our present case, the inclusive production cross sections $\sigma(D^{\pm})$ 
and $\sigma(W^{\pm})$ have been calculated in accordance with Refs. \cite{JKLZ} 
and \cite{inclW}, respectively. For the considered fiducial phase space our 
expectations read
\begin{eqnarray}
\sigma_{\mbox{incl}}(D^+)=\sigma_{\mbox{incl}}(D^-)&{=}&11.4\;\mu\mbox{b},\\   
\bW\sigma_{\mbox{incl}}(W^+)&{=}&3.5\;\mbox{nb},\label{wp}\\
\bW\sigma_{\mbox{incl}}(W^-)&{=}&2.5\;\mbox{nb},\label{wm}
\end{eqnarray}
where the estimates (\ref{wp})-(\ref{wm}) are supported by direct recent
measurement \cite{ATLASW}, and so,
\begin{eqnarray}
\bW\sigma_{DPS}(W^+D^+) &=& 2.7 \mbox{~pb}\\
\bW\sigma_{DPS}(W^-D^-) &=& 1.9 \mbox{~pb}.
\end{eqnarray}

These numbers are close to the combined SPS contribution. This means that
the excess brought by DPS to the visible $W^\pm D^\pm$ cross-sections is 
not large enough to unambiguously testify for its presence: the DPS signal 
is large, but the background uncertainties are also large. Moreover, the 
shapes of the DPS and SPS kinematic distributions are rather similar:
the decays of heavy $t$-quarks and $W$ bosons
make the final state distributions broad and smooth.
Selection cuts on the azimuthal angle difference $\Delta\phi$ 
or rapidity difference $\Delta y$, so promising in other reactions 
\cite{JJ_DPS}, remain practically useless in the present case.
This is illustrated in Fig. 1 where we show correlations between $D^\pm$
mesons and muons coming from $W^\pm$ bosons in same-sign events at ATLAS
conditions.

Fortunately, the indirect contributions can be significantly reduced (if not
rejected completely) using a well-known experimental technique based on the
property that the secondary $b$-decay vertex is displaced with respect to 
the primary interaction vertex. We are then left with direct background 
(\ref{s1})-(\ref{s2q}) lying well below the DPS level,
even with conservative estimate of $\sigma_{\rm eff}{=}15$ mb and with 
`pessimistic' choice of $\mu_R$ as dicussed above. In fact, our numbers
represent the upper edge of the background uncertainty band. A similar 
relation is seen for $D^*$ mesons:
\begin{eqnarray}
&\bW\sigma_{DPS}(W^+D^{*+}) = 2.3 \mbox{~pb}&\\
&\bW\sigma_{DPS}(W^-D^{*-}) = 1.6 \mbox{~pb}&\\
&\bW\sigma^{W^+D^{*+}}(u\bar{d}{\to}Wc\bar{c})= 0.35 \mbox{~pb}& \label{s1a}\\
&\bW\sigma^{W^-D^{*-}}(d\bar{u}{\to}Wc\bar{c})= 0.25 \mbox{~pb}&\label{s2a}\\
&\bW\sigma^{W^+D^{*+}}(gu{\to}Wdc\bar{c})= 0.85 \mbox{~pb}& \label{s1qa}\\
&\bW\sigma^{W^-D^{*-}}(gd{\to}Wuc\bar{c})= 0.60 \mbox{~pb}&\label{s2aq} 
\end{eqnarray}

We thus come to an important conclusion that the production of same-sign
$W^\pm D^{*\pm}$ states is very indicative as DPS signal. This situation is 
close to the production of same-sign $W^\pm W^\pm$ pairs proposed earlier 
in Ref. \cite{Kulesza}. However, the $W^\pm W^\pm$ events occur at a much 
lower rate and are then less convenient for analysis.

\section{Conclusions}

We have considered the production of a W boson in association with a charmed
meson in $pp$ collisions at the LHC and made a comparison with experimental 
results. Our theoretical calculations have shown reasonable agreement with 
ATLAS data on $\sigma^{OS-SS}(WD)$. We have extended our consideration to 
the same-sign $W^\pm D^\pm$ configurations and found that after rejecting 
the $b$-decays the DPS signal clearly dominates over SPS background.
Thus, we come to an important conclusion that the production of same-sign 
$W^\pm D^\pm$ states can serve as a new reliable indicator of double parton
scattering.

\section*{Acknowledgments}
The authors thank A.V. Berezhnoy for useful discussions.
This work is supported in part by the Federal Agency for Science and 
Innovations of Russian Federation (Grant NS-3042.2014.2)
and by the DESY Directorate in the framework of Moscow-DESY project
on Monte-Carlo implementations for HERA-LHC.


\begin{figure}
\begin{center}
\epsfig{figure=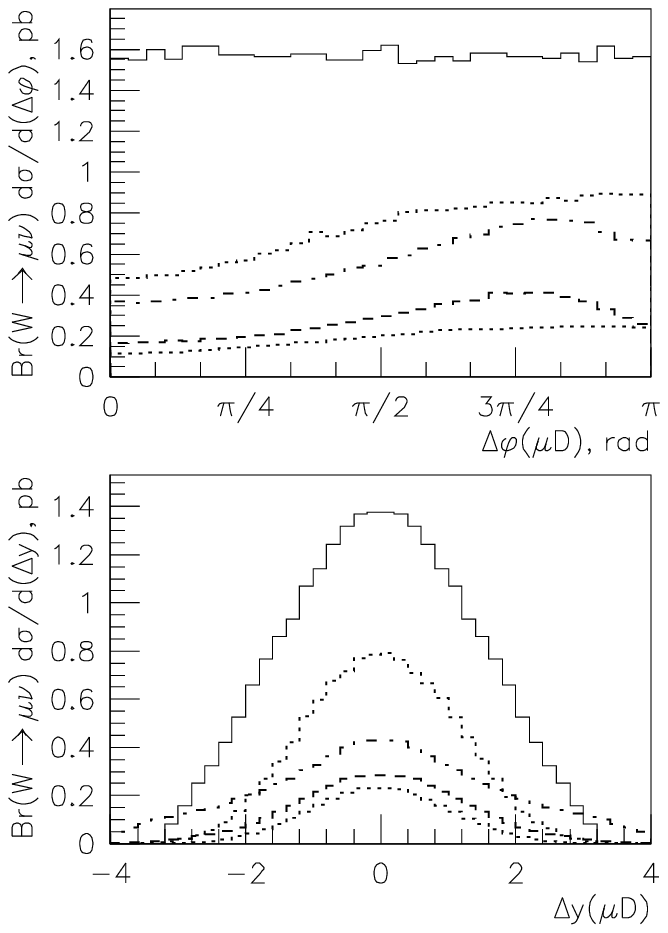,width=8.5cm}
\caption{Kinematic correlations between muons and $D$-mesons
in same-sign events ($\mu^\pm D^\pm$) at ATLAS conditions: 
distributions in the azimuthal angle difference $\Delta\phi$ (upper panel) 
and rapidity difference $\Delta y$ (lower panel).
The different contributions are represented by:
solid curve, double parton scattering;
upper dotted curve, $gg{\to}t\bar{t}$;
lower dotted curve, $q\bar{q}{\to}t\bar{t}$;
dashed curve, $u\bar{d}{\to}Wc\bar{c}$ and $d\bar{u}{\to}Wc\bar{c}$;
dash-dotted curve, $gu{\to}Wdc\bar{c}$ and $gd{\to}Wuc\bar{c}$.
}
\end{center}
\end{figure}

\begin{thebibliography}{99}
\bibitem{Bartalini} P. Bartalini \etal, arXiv:1111.0469.
\bibitem {Abramowicz} H. Abramowicz \etal,  arXiv:1306.5413.
\bibitem{Bansal} S. Bansal \etal, arXiv:1410.6664.
\bibitem{ATLAS} G. Aad \etal (ATLAS Collab.), \jhep{05}{068}{2014} .
\bibitem{GLR}
 L.V.Gribov, E.M.Levin, M.G.Ryskin, \prep{100}{1}{1983} ;\\
 E.M. Levin, M.G. Ryskin, Yu.M. Shabelsky, A.G. Shuvaev,
                                       \sjnp{53}{657}{1991} ;\\
 S. Catani, M. Ciafaloni, F. Hautmann,
                       \plb{242}{97}{1990} ; \np{B366}{135}{1991} ;\\
 J.C. Collins, R.K. Ellis, \np{B360}{3}{1991} .
\bibitem{smallX}
 B. Andersson \etal (Small x Collab.), \epjc{25}{77}{2002} ;\\
 J. Andersen \etal (Small x Collab.), \epjc{35}{67}{2004} ;\\
 J. Andersen \etal (Small x Collab.), \epjc{48}{53}{2006} .
\bibitem{KMR} M.A. Kimber, A.D. Martin, M.G. Ryskin
                                             \prd{63}{114027}{2001} .
\bibitem{MSTW} A.D. Martin, W.J. Stirling, R. S. Thorne, G. Watt,
                                             \epjc{63}{189}{2009} .
\bibitem{Peterson} C. Peterson, 
      D. Schlatter, I. Schmitt, P.M. Zerwas, \prd{27}{105}{1983} .
\bibitem{JKLZ} H. Jung, M. Kraemer, A.V. Lipatov, N.P. Zotov,
                              \jhep{01}{085}{2011} .
\bibitem{PDG} K.A. Olive \etal (Particle Data Group),
                                            \chp{C38}{090001}{2014} .
\bibitem{cdf4j} F. Abe \etal (CDF Collab.), \prd{47}{4857}{1993} .
\bibitem{cdf}   F. Abe \etal (CDF Collab.), \prd{56}{3811}{1997} .
\bibitem{d0}  V.M. Abazov \etal (D0 Collab.), \prd{81}{052012}{2010} .
\bibitem{atlas} G. Aad \etal (ATLAS Collab.), \njp{15}{033038}{2013} .
\bibitem{cms}   S. Chatrchyan \etal (CMS Collab.), \jhep{03}{032}{2014} .
\bibitem{D0}  V.M. Abazov \etal (D0 Collab.), \prd{90}{111101}{2014} .
\bibitem{inclW} 
      S.P. Baranov, A.V. Lipatov, N.P. Zotov, \prd{78}{014025}{2008} .
\bibitem{ATLASW} G. Aad \etal (ATLAS Collab.), \prd{85}{072004}{2014} .
\bibitem{JJ_DPS} S.P. Baranov, W. Sch\"{a}fer, A. M. Snigirev,
                 A. Szczurek, N.P. Zotov, \prd{87}{034035}{2013} .
\bibitem{Kulesza} A. Kulesza, W.J. Striling, \plb{475}{168}{2000} .
\end{thebibliography}
\end{document}